\documentclass[aps,prl,preprintnumbers,amsmath,amssymb,superscriptaddress,twocolumn,10pt]{revtex4-2}
\usepackage{standalone}
\usepackage{txfonts}
\usepackage{graphicx}
\usepackage{dcolumn}
\usepackage{bm}
\usepackage{array}
\usepackage[dvipsnames]{xcolor}
\usepackage[%
    colorlinks=true,
    pdfborder={0 0 0},
    linkcolor=red
]{hyperref}
\usepackage{chngpage}
\usepackage{appendix}
\usepackage{xspace}
\usepackage{comment}

\graphicspath{{Figures/}}

\newcommand{\supplementarysection}{%
  \setcounter{figure}{0}
  \let\oldthefigure\thefigure
  \renewcommand{\thefigure}{S\oldthefigure}
  \setcounter{table}{0}
  \let\oldthetable\thetable
  \renewcommand{\thetable}{S\oldthetable}
}

\begin{document}

\def\rvs {RbV$_3$Sb$_5$\xspace}
\def\cvs {CsV$_3$Sb$_5$\xspace}
\def\kvs {KV$_3$Sb$_5$\xspace}
\def\V {$^{51}$V\xspace}
\def\Sb {$^{121}$Sb\xspace}
\def\Rb {$^{87}$Rb\xspace}
\def\usr {$\mu$SR\xspace}
\def\zfusr {ZF-$\mu$SR\xspace}

\title{Unveiling the nature of electronic transitions in RbV$_3$Sb$_5$ with Avoided Level Crossing $\mu$SR}

\author{Pietro Bonf\`a}
\affiliation{Dipartimento di Fisica, Informatica e Matematica, Universit\`a di Modena e Reggio Emilia, Via Campi 213/a, 41125 Modena, Italy}

\author{Francis Pratt}
\affiliation{ISIS Pulsed Neutron and Muon Source, Rutherford Appleton Laboratory, Didcot OX11 0QX, U.K.}

\author{Diego Valenti}
\affiliation{
Dipartimento di Scienze Matematiche, Fisiche e Informatiche, Universit\`a di Parma, I-43124 Parma, Italy
}

\author{Ifeanyi John Onuorah}
\affiliation{
Dipartimento di Scienze Matematiche, Fisiche e Informatiche, Universit\`a di Parma, I-43124 Parma, Italy
}

\author{Anshu Kataria}
\affiliation{
Dipartimento di Scienze Matematiche, Fisiche e Informatiche, Universit\`a di Parma, I-43124 Parma, Italy
}

\author{Peter J. Baker}
\affiliation{ISIS Pulsed Neutron and Muon Source, Rutherford Appleton Laboratory, Didcot OX11 0QX, U.K.}

\author{Stephen Cottrell}
\affiliation{ISIS Pulsed Neutron and Muon Source, Rutherford Appleton Laboratory, Didcot OX11 0QX, U.K.} 

\author{Andrea Capa Salinas} 
\affiliation{
Materials Department, University of California Santa Barbara,
Santa Barbara, California 93106, USA}

\author{Stephen D. Wilson}
\affiliation{
Materials Department, University of California Santa Barbara,
Santa Barbara, California 93106, USA}

\author{Zurab Guguchia}
\affiliation{PSI Center for Neutron and Muon Sciences CNM, 5232 Villigen PSI, Switzerland}

\author{Samuele Sanna}
 \affiliation{
Dipartimento di Fisica e Astronomia ``A. Righi'', Universit\`a di Bologna, I-40127 Bologna, Italy }

\date{\today}

\begin{abstract}

Kagome superconductors AV$_{3}$Sb$_{5}$ provide a unique platform for studying the interplay between a variety of electronic orders, including superconductivity, charge density waves, nematic phases and more.
Understanding the evolution of the electronic state from the charge density wave to the superconducting transition is essential for unraveling the interplay of charge, spin, and lattice degrees of freedom giving rise to the unusual magnetic properties of these nonmagnetic metals.
Previous zero-field and high-field \usr studies revealed two anomalies in the muon spin relaxation rate, a first change at $T_{CDW} \sim 100$~K and a second steep increase at $T^{*}\sim 40$~K, further enhanced by an applied magnetic field, thus suggesting a contribution of magnetic origin.
In this study, we use the avoided level crossing \usr technique to investigate charge order in near-zero applied field. By tracking the temperature dependence of quadrupolar level-crossing resonances, we examined the evolution of the electric field gradient at V nuclei in the kagome plane. Our results show a significant rearrangement of the charge density starting at $T^{*}$ indicating a transition in the charge distribution, likely electronic in origin, well below $T_{CDW}$. These findings, combined with previous \usr, STM, and NMR studies, emphasize the intertwined nature of proximate phases in these systems, with the charge rearrangement dominating the additional increase in \usr relaxation rate below $T^{*}$.

\end{abstract}

\maketitle

\section{Introduction}
Transition-metal based kagome materials AV$_3$Sb$_5$ (A = K, Rb, Cs) have generated increasing interest in the scientific community owing to the diverse physical properties that have been observed, including nontrivial band topology, anomalous Hall effect (AHE) \cite{Yang_2020}, and an intriguing interplay between superconductivity (SC) and unconventional charge density wave (CDW) \cite{Frachet_2024}.

Despite numerous theoretical and experimental investigations, even when focusing only on the normal state, the current understanding of the cascade of transitions characterizing the electronic behavior of these compounds remains incomplete.
Initially, with decreasing temperature, a CDW transition occurs at $T_{CDW} \sim 78 - 102$~K depending on the alkali metal \cite{Wenzel_2022}. In this phase, the hexagonal lattice with $P6/mmm$ symmetry undergoes a distortion involving the formation of V hexamers and trimers which produces the so called Tri-Hexagonal structure (TrH) in the kagome planes with the crystal adopting the $Fmmm$, $Cmmm$ or $C2/m$ space group symmetry \cite{Li_2021, Subedi_2022, Tan_2021, Frassineti_2023, Linus_2023,Ptok_2022}.
This phase also features an additional modulation along the $c$ axis, likely due to 
a staggered displacement pattern of the kagome layers, although some uncertainty on the mutual arrangement of these planes persists \cite{Ortiz_2021, Stahl_2022, Zhang_2024, Li_2021, Scagnoli_2024}, probably owing to the delicate competition between CDW orders with different stacking modulations \cite{Park_2023}.

At lower temperature, the presence of a nematic transition breaking the $C_6$ symmetry of the kagome planes
\footnote{It should be noted that, while at $T_{CDW}$ the rotational symmetry of the system changes form $C_6$ to $C_2$ owing to the staggered order along $c$, the kagome layers retain the original sixfold rotational symmetry. Yet, at lower temperature, the $C_6$ symmetry of the 2D kagome layer may also be broken by the so called ``strong'' nematic state (to distinguish it from the ``weak'' nematic state realized at $T_{\text{CDW}}$
\cite{Grandi_2023}).
}
has been reported for \cvs \cite{Zhao_2021}, but its nature (and existence) is still
debated \cite{Nie_2022, Sur_2023, Li_2023, Liu_2024, Asaba_2024, Frachet_2024}.
Notably, this and other symmetry-breaking charge orders were identified only in \cvs \cite{Wilson_2024}, while no other charge order transitions between $T_{\text{CDW}}$ and $T_{\text{SC}}$ have been reported for A=K, Rb.

A distinctive feature of kagome systems is the potential for time-reversal symmetry (TRS) breaking in the normal state, as initially suggested by high-field scanning tunneling microscopy \cite{Jiang_2021}, along with evidence from a combination of zero-field and high-field muon spin relaxation (\usr) \cite{Mielke_2022}, supported by various experiments \cite{Chen_2022,Guo_2022,Xing_2024, deng2024evidencechaintimereversalsymmetrybreaking}
such as magneto-chiral anisotropy, the anomalous Hall effect\cite{Yang_2020}, and magneto-optical Kerr effect (MOKE) measurements \cite{Xu_2022}. 

Several theoretical studies \cite{Lin_2021,Park_2021,Feng_2021,Denner_2021,Christensen_2022, Li_2024,Shimura_2024,Dong_2023,Tazai_2023} attribute the TRS-breaking signal to orbital current phases. Furthermore, polarized neutron diffraction experiments hint at the presence of a weak magnetic signal in the second Brillouin zone at $M_2 = (1/2, 1/2, 0)$ \cite{liege_2024}. This finding, interpreted as loop current patterns localized on vanadium triangles, reports an ordered orbital magnetic moment of at most 0.02 $\mu_{B}$ per vanadium triangle. 

However, whether the system
breaks TRS spontaneously, i.e.
in zero applied magnetic field, has been challenged by MOKE \cite{Saykin_2023,Wang_2024,Farhang_2023} studies that have reported no observable Kerr response in zero field (ZF).
A recent analysis of transport properties suggests that the origin of the controversial reports is an extraordinary sensitivity to weak perturbations, in particular strain and magnetic fields \cite{Guo_2024}.
This is in line with earlier high-field \usr experiments revealing a significant enhancement of muon relaxation rates with out-of-plane magnetic fields, and with high-field STM experiments, which showed charge density wave (CDW) intensity switching under out-of-plane magnetic fields and in-plane electric fields, implying an unusual piezo-magnetic response \cite{Xing_2024}.

\zfusr is one of the cleanest methods to identify TRSB owing to the possibility of  performing ZF experiments on bulk samples and thanks to its sensitivity to extremely small magnetic fields \cite{Blundell_2022}. For this reason, a large number of experiments have been conducted \cite{Yu_2021,Khasanov_2022}.
The central quantity investigated during the experiment is the polarization function of the muon, defined as the time evolution of the muon spin projected along the direction of the positron detectors, which is obtained by exploiting the asymmetry in the weak decay of the muon.
This time evolution encodes the details of the interaction between the muon spin, the neighboring nuclear spins and electronic (spin or orbital) moments, making it a local probe for both electronic and structural transitions.

Previous ZF  \usr studies \cite{Mielke_2022, Khasanov_2022, Guguchia_2023, Wilson_2024} on \rvs have revealed a two-step increase in the relaxation rate, a smaller one at $T_{CDW} \simeq 100$ K and a second one at $T^{*} \lesssim 50$ K. The increase in the relaxation rate corresponds to internal fields on the order of 0.01~mT. The effect is enhanced under an applied magnetic field, thus suggesting a magnetic contribution to the relaxation. 
A significant enhancement of the relaxation is also detected near the surface region of RVS, specifically within 30 nm from the surface. A similar two-step increase in the relaxation rate was observed in the sister compound \cvs, at $T_{CDW} \simeq 90$ K and at $T^{*} \simeq 30$ K. 
However, the nature of the additional increase in the relaxation rate below $T^{*}$ remains uncertain, leaving open the question of whether this increase is magnetic in origin or if changes in the charge order also contributes.

Muons serve as exceptional probes not only for magnetism but also for investigating charge order. In particular, the experimental technique known as Avoided Level Crossing (ALC) \usr enables the study of charge distribution evolution by monitoring changes in the Electric Field Gradient (EFG) tensor at nuclei with spin $I>1/2$, located near the muon. This information is obtained indirectly through the dipolar coupling between the muon and quadrupole-active nuclei.
The idea behind ALC was first proposed by Abragam \cite{Abragam_1984}, who noted that nuclear quadrupole splittings could be measured indirectly with \usr, by finely tuning the muon Zeeman energy to
match a nuclear quadrupolar splitting, thus resulting in a cross relaxation between the spins known as muon quadrupole level crossing resonance ($\mu$-QLCR or simply QLCR).
As a consequence, one can selectively probe different nuclei by matching different quadrupolar energies,
whereas in ZF experiments the nuclear contribution to the polarization function is dominated by the dipolar interaction between the muon and the nearest nuclei.
This is indeed the strategy adopted in this work and a nice review of this approach is presented in Ref.~\cite{Cox_1992}.

\begin{figure}
\includegraphics{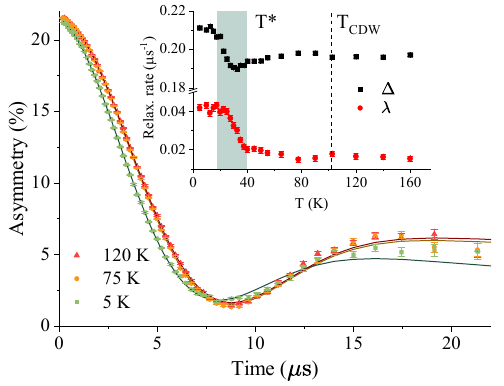}
\caption{The \zfusr spectra of \rvs. In the main panel, three high-statistics acquisitions showing the evolution of the asymmetry as a function of the temperature. A tiny difference is observed between signals above (120 K) and below (75 K) the CDW transition (102 K).
A noticeably faster relaxation is instead observed in the measurement performed at 5~K.
The continuous lines are best-fits to Eq.~\ref{eq:zf}.
The inset shows the temperature evolution of the fitting parameters resulting from the analysis described in the main text. \label{fig:ZF}}
\end{figure}

\section{Results}
For all experiments we used finely ground \rvs powders, identical to the sample in Ref.~\cite{Graham_2024}. 
Both the ZF and the ALC acquisitions have been carried out using the EMU spectrometer \cite{Giblin_2014} at the ISIS pulsed muon facility.

\begin{figure*}
\includegraphics{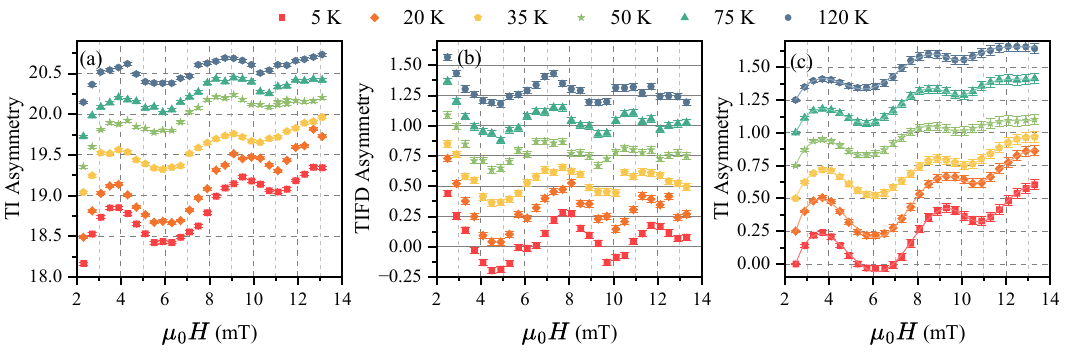}
\caption{ALC results and analysis. In panel (a), the time-integrated asymmetry as a function of the temperature. Results are reported with vertical offsets for clarity. In (b), the time-integrated-field-differential acquisitions. The same vertical offset is used. In (c) the numerical integration of the results shown in panel (b). The continuous lines are best fit to the data according to Eq.~\ref{eq:alcfit} described in the main text.\label{fig:alc}}
\end{figure*}

\subsection{Zero Field \usr}

The \zfusr data are shown in Fig.~\ref{fig:ZF}. The asymmetry has been analyzed with best fits to the equation
\begin{equation}
    A(t) = A_{0} \left[ \frac{1}{3} + \frac{2}{3} \left( 1 - \Delta^2 t^2\right) e^{-\Delta^2 t^2 / 2}  \right] e^{-\lambda t}  + B\label{eq:zf}
\end{equation}
for consistency to the previous literature \cite{Guguchia_2023,Shan_2022,Khasanov_2022}.
In Eq.~\ref{eq:zf}, $A_{0}$ is the relaxing asymmetry, $B$ is a baseline, and the function in square brackets is the Kubo-Toyabe (KT) function \cite{Blundell_2022} characterized by a relaxation rate $\Delta$ and multiplied by an additional Lorentzian relaxation, parameterized by $\lambda$.
This phenomenological approach matches very well with the experimental data at short time, while small deviations are observed for $t>13$~$\mu$s in the $T=75$ and 120~K measurements.
A dramatic discrepancy is instead observed at 5~K for $t > 10\, \mu$s, showing the limits of the phenomenological description based on Eq.~\ref{eq:zf}.
Notably, the tail of the 5~K signal remains relatively flat, indicating a lack of dynamic effects.

It should be noted that the KT function in Eq.~\ref{eq:zf} arises from a semi-classical description of the muon-nuclei interaction. For this reason, an additional relaxation contribution and a (temperature-dependent) baseline $B$ are required.
Departures from the semi-classical prediction are common and are generally found in the long-time tail, as shown for example in Ref.~\cite{Huang_2012}, though relevant effects can also take place at shorter times \cite{Bonfa_2022,Wilkinson_2023}.
A very accurate prediction of the ZF muon polarization function above $T^{*}$ can be obtained from first principles modeling of the muon site and its interaction with the neighboring nuclei, when the entire description is performed at the quantum mechanical level. These results have already been presented in Ref.~\cite{Graham_2024}, together with a detailed description of the muon sites, which reveals that the ZF signal is most sensitive to the in-plane Sb atoms through a simple dipolar interaction between the muon spin and the nearest neighbouring isotopes ($m_{^{121}Sb} = 3.36 \mu_P$, $m_{^{123}Sb} = 2.55 \mu_P$ and $d_{\mu-Sb} \sim 1.7$~\AA) Sb isotopes, while the coupling with the V nuclei forming the kagome lattice is much weaker ($m_{^{51}V} = 5.15 \mu_P$ and $d_{\mu-V} \sim 3.5$~\AA).
However, in the following discussion, we still opt for the phenomenological model due to its simplicity and effectiveness in capturing the transitions observed in the ZF data.

The inset of Fig.~\ref{fig:ZF} reports the temperature evolution of $\lambda$ and $\Delta$ obtained from  best-fits of the measurements performed in the temperature interval $5 - 160$~K.
Two transitions can be detected, at about $T_{\text{CDW}}=100$~K and $T^{*}=40$~K. The trend follows what has been already reported in literature \cite{Guguchia_2023}.
As can be appreciated from the main panel Fig.~\ref{fig:ZF}, the transition at $T_{\text{CDW}}$ has a small effect on the polarization function of the muon. On the other hand, a marked change takes place below $T^{*}$ as shown also by the difference between the raw data acquired at 5 and 75~K (Fig.~\ref{fig:ZF} main panel).

The extended time window of our new measurements provides a more detailed picture of the evolution of the \usr signal as a function of the temperature.
From the phenomenological analysis, considering both $\lambda$ and $\Delta$, the broad transition that starts at $T^{*}=40$~K is finally completed at $T \sim 20$~K, where both the relaxation coefficients become temperature independent.

\subsection{Avoided Level Crossing}
Longitudinal field ALC experiments have been performed with $\mu_{0}H$ ranging from 2.3 to 13.1~mT.
The asymmetry within the time interval 1.44~$\mu$s to 30~$\mu$s is integrated in time (TI) and shown in Fig.~\ref{fig:alc}a.
For $\mathbf B \to 0$ the signal decays owing to the dipolar interaction between the muon and the neighbouring nuclei and the TI asymmetry reduces accordingly.
With increasing longitudinal fields, two resonances at about 6 and 11~mT are clearly observed, especially at low temperature.

In order to reduce the noise of the data, we employ a field-differential approach \cite{Kadono_2000},
where the external magnetic field is varied by $\pm 0.2$~mT during the acquisition, thus removing the instabilities of the beam affecting the raw data. The difference of the asymmetries acquired in the two conditions represents the finite difference approximation of the derivative of the signal as a function of the applied magnetic field \cite{Kadono_2000}.
The time-integrated field-differential (TIFD) signal is reported in Fig.~\ref{fig:alc}b.
These curves already show a clear temperature dependence, but the main result of this procedure are the resonances shown in Fig.~\ref{fig:alc}c obtained by numerical integration.

The numerically integrated TIFD data, $I(B)$, are fitted to the equation
\begin{equation}
    I(B) = I_{0} \left(1-\frac{\tau}{B^{N}}\right) - \sum_{i=1}^{2} 
    \frac {A_{i}}{\sigma_{i} {\sqrt {2\pi }}}\exp \left(-{\frac {1}{2}}\frac {(B-B^{\mathrm{res}}_{i} )^{2}}{\sigma_{i} ^{2}}\right) \label{eq:alcfit}
\end{equation}

where in the first two terms, $\tau$ and $N$, are used to approximate the exact description \cite{Yaouanc_2011} of the time integrated polarization function of pure dipolar origin in longitudinal applied field.
The observed level-crossing resonances are instead captured with two Gaussian functions, each characterized by three parameters: $A$, $B^{\mathrm{res}}$ and $\sigma$ respectively the area, the average resonance field and the width of the resonance.

\begin{figure*}
\includegraphics{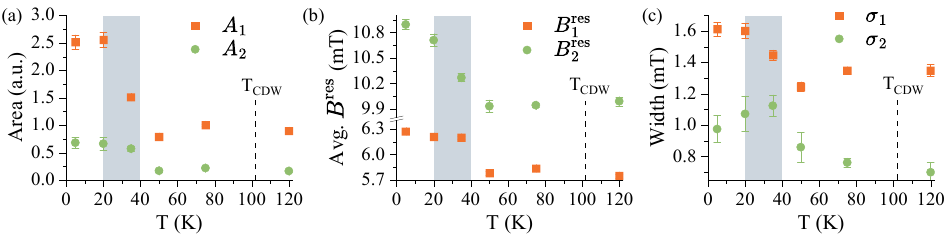}
\caption{Parameters obtained from best-fit curves shown in Fig.~\ref{fig:alc}c.
In panel (a) the area of the Gaussian used to describe the two resonances. In (b) the resonant field, in (c) the width of the two Gaussian functions.
The gray shadow shows the same temperature interval displayed in Fig.~\ref{fig:ZF} where the second low-temperature transition takes place according to ZF results.\label{fig:summary}}
\end{figure*}

The results of the fit are shown in Fig.~\ref{fig:summary}.
Surprisingly, the CDW transition
has a very slight effect, if any, on the resonance parameters, while sizable deviations are observed for $T \lesssim T^*$.
Indeed the area of both resonances, shown in Fig.~\ref{fig:summary}a, increases below 40~K and it remains constant between 20 and 5~K. The average resonant fields, shown in Fig.~\ref{fig:summary}b, also display a clear shift of the order of 1~mT when the temperature drops below 40~K.
Finally, a similar temperature dependence is also observed for the width of the resonance peaks in panel (c), although the actual trend for the second resonance is impaired by the large uncertainty of this parameter.

\begin{figure}
\includegraphics{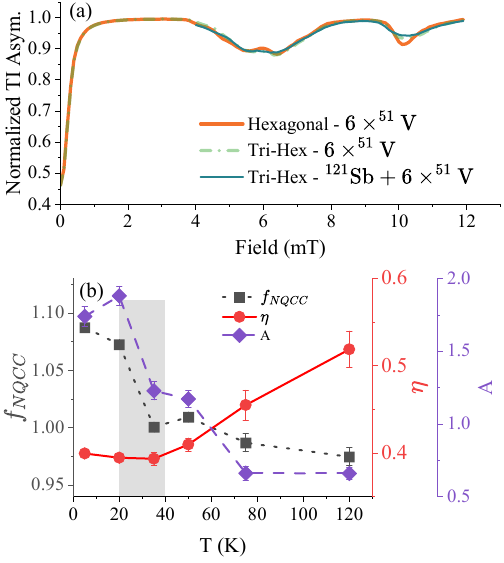}
\caption{Panel (a): predicted resonances for $T>T_{CDW}$ in the Hexagonal structure and for the low temperate TrH structure (neglecting the modulation along the $c$ axis). The legend also reports the cluster of nuclei included in the simulation. The EFGs at V and Sb sites are obtained from a full-potential DFT-based prediction.
Panel (b): results of the fit of TI asymmetry to the microscopic model described in the main text.\label{fig:theory}}
\end{figure}

\section{Discussion}

In order to obtain a microscopic description of the ALC results we started from the Plane Wave (PW) based DFT simulations published in Refs.~\cite{Graham_2024, materialscloud}.
As already discussed in \cite{Graham_2024}, in the staggered TrH structure there are 3 symmetry inequivalent muon sites, but the neighborhoods of the muon are always very similar, i.e. the distance between the muon and the various atoms of the system change very little between the sites.
These simulations also provide information on the perturbation introduced by the muon on the EFG at the neighboring sites.
As expected, we find that the EFG of the nearest neighbor (NN) of the muon, the in-plane Sb atom, is strongly perturbed by the presence of the positive interstitial charge.
In contrast, the EFG at the second NN, the Rb atom, is much less affected, with variations of $V_{zz}$ smaller than 10\% and even less for the V atoms that are more than 3.5~\AA{} away from the muon (see Supplemental Material for details).
This preliminary analysis provides valuable information towards the identification of the nuclei involved in a QLCR producing the experimental signal.
It is found indeed that, for the field range of the measurement, a hexagon of six V atoms close to the muon site contribute most significantly to the QLCR spectra (see SM).

While PW based results represent a good starting point, their accuracy is lower than full-potential APW based simulations in the description of charge distributions at nuclear sites.
In light of this, to obtain more accurate predictions, we adopt the Full-Potential APW-based estimate for the EFGs at V atoms in the TrH phase already published in Refs.~\cite{Frassineti_2023, Frassineti_materialscloud} while maintaining the equilibrium distances between the muon and the atoms obtained with the PW-based supercell simulation \cite{Graham_2024, materialscloud}.
By doing so we neglect the staggering along the $c$ axis and the perturbation introduced by the muon on the EFG (which is very limited for V atoms, see SM). In the TrH structure there are only two symmetrically distinct sites and we compute, for each of them, the polarization function arising from the interaction between the muon and six NN V atoms, using the method introduced by Celio \cite{Celio_1986, Bonfa_2021}.
To check for convergence of our simulations, we also include the NN Sb atom (the $^{121}$Sb isotope is only considered).
The nuclear quadrupolar coupling constant $C_Q$ , defined as 
\begin{equation}
     {C}_Q={e}_{0}V_{zz}Q/h \label{eq:fp}
\end{equation}
where $Q=-0.043$ barn is the quadrupolar moment of $^{51}$V and $V_{zz}$ is the largest eigenvalue of the EFG tensor, is predicted to be $C_{Q}^{APW}=6.85$~MHz, and the asymmetry parameter $\eta = (V_{xx} - V{yy})/V_{zz}$ is $\eta ^{APW} = 0.43$.
The inclusion of the NN Sb in the Hilbert space negligibly affects the resonances and only alters the TI Asymmetry for $B\to 0$ (not shown) \footnote{This is trivially the result of the larger dipolar contribution of the closest Sb nucleus with respect to the more distant V nuclei}.

The resulting curves, shown in Fig.~\ref{fig:theory}a, 
match very well with the experimental results above $T^*$. 
In addition, they reveal why this probe is insensitive to the CDW order: the EFG at V nuclei varies very little across the transition and the geometrically equivalent V atoms in the high temperature phase split into two inequivalent sites with equal multiplicity. For the first(second) set, $V_{zz}$ is slightly decreasing(increasing) with respect to the value in the hexagonal phase. The resulting signal displays QLCR that are just very slightly broader than the ones of the hexagonal phase, with the only noticeable difference visible at about 11~mT.

Unfortunately, this approach is computationally demanding and prevents fitting the microscopic QLCR parameters $C_Q$ and $\eta$ to the experimental data.
Therefore, since the muon is close to the centre of the hexagon and the magnitudes of the EFG parameters are similar for all six, for fitting purposes we took one V site as representative and calculated the polycrystalline averaged QLCR spectrum versus $\eta$ with the principal axis of the field gradient aligned at 90$^\circ$ to the muon vector. 
The QLCR for the six V sites was then evaluated by scaling the resonance amplitude by a factor of six. 
Finally, a cubic background is added to account for the minor contributions from other nuclei. 
This approach yields an average trend and assumes that the dipolar interactions between V nuclei are not significant compared to the dipolar interaction between the muon and each individual V. This model requires a reduced Hilbert space (consisting of the muon and one V atom at a time), allowing us to extract an averaged trend for the parameters describing the dipolar and quadrupolar coupling of the V nuclei with the muon and the surrounding electronic charge, respectively.

In order to obtain a quantitative comparison with the experiment, we fit the time integrated asymmetry (instead of the numerical integration of Fig~\ref{fig:alc}c) and we report the results with respect to the \textit{ab initio} prediction.
The comparison is made through the NQCC factor, defined as $f_{NQCC} = {C}_Q ^{\text{Exp}}/{C}_Q ^{\text{PW}}$ and the scaling factor $A$ required to match the amplitude of the experimental resonance with the predicted one. The asymmetry parameter, $\eta$, is instead reported directly for simplicity of presentation \footnote{For the sake of consistency, we adopt the prediction of PW simulations for all parameters definig the interaction i.e. the distance between the muon and V atoms and the EFG at V sites. The latter values are averaged among the six nuclei and $C_{Q}^{\text{PW}}$ and $\eta^{\text{PW}}$ are 6.893~MHz and 0.164, respectively. The prediction for $V_{zz}$ is very close to the one from APW simulations in the unperturbed lattice while the one for $\eta$ deviates substantially: this is actually due to the limits of the computational basis rather than to perturbation effects. Indeed, unperturbed V nuclei far from the muon have $\eta \sim 0.2$ instead of $\eta \sim 0.4$. See supplemental materials for more details.}.

The results of the fit are shown in Fig.~\ref{fig:theory}b.
At high temperature, the resonance frequency is very well estimated ($f_{NQCC} \sim 0.97$) while the amplitude of the resonance is slightly under-estimated ($A \sim 0.75$).
The asymmetry parameter $\eta$ is close to the experimental result \cite{Zhang_2024} and to APW based estimates.
Both $A$ and $f_{NQCC}$ increase as temperature decreases, and the trend of the microscopic parameters roughly aligns with the phenomenological analysis in Fig.\ref{fig:summary}.
Interestingly, the variation of $V_{zz}$ at $V$ nuclei probed by muons, of the order of 5\%, is slightly larger than the one reported by high field NMR measurements in \cvs (see SI of Ref.~\cite{Nie_2022}).

With the microscopic origin of the QLCR clarified, we now turn our attention to the relationship between these findings and previous zero-field (ZF) measurements.

Remarkably, the ZF muon-spin relaxation rates $\Delta$ and $\sigma$ (inset of Fig.~\ref{fig:ZF}), the phenomenological ALC parameters $A_{i}$, $B_{i}$, and $\sigma_{i}$ (Fig.~\ref{fig:summary}), and the results of the microscopic model (Fig.~\ref{fig:theory}b) exhibit similar temperature dependence, as highlighted by the shaded region. A key question is whether the ALC resonance field shift, shown in Fig.~\ref{fig:summary}b, can be attributed to an additional field generated by weak magnetic interactions. The ZF relaxation rate increasing by $\delta \lambda \lesssim 0.03$~$\mu \text{s}^{-1}$, corresponds to a local field increase of $\delta B = \delta \lambda / (\gamma_{\mu}) \sim 0.035$~mT \footnote{This value is computed by considering the half width half maximum of a Lorentzian distribution. A similar conclusion is also reached by considering the variation of the $\Delta$ parameter.}. However, the observed shift in ALC resonance is more than an order of magnitude larger. Therefore, the contribution to the increase of ALC resonance field from magnetic coupling between the muon and the electronic channel via orbital or spin magnetism is minimal. We conclude that the increase of the resonant field is primarily due to a shift in nuclear quadrupolar energy levels, caused by charge redistribution at the V sites, indicating the presence of a transition in the charge channel setting in before the onset of superconductivity.

The key question now is what drives this charge redistribution. Given the lack of thermodynamic evidence for a transition at $T^{*} \sim 40$ K and the absence of major structural distortions, this cannot be attributed to simple structural changes. Notably, NMR experiments and STM measurements in \cvs and \kvs \cite{Li_2023, Nie_2022,Zhang_2024c} have reported rotational symmetry broken states stabilized well below the CDW order.
It is plausible to assume that the charge redistribution observed below $T^{*} \sim 40$ K in \rvs shares a similar origin with \cvs and \kvs hinting at electronic nematicity.

\section{Conclusions}

We have presented ZF and ALC \usr experiments performed on \rvs. 
With the latter technique we investigated the evolution of charge order in the kagome plane. Our results reveal a significant rearrangement of charge density around the muon below $T^{*} \sim 40$ K, an effect that matches with the upturn of the muon relaxation rate observed in \zfusr. This uncovers a hitherto unnoticed charge order transition well below the onset of the CDW and before the system enters the superconducting state and suggests that a non-trivial evolution of the local charge/electronic landscape is the primary origin for the phenomenology observed in zero and near-zero field conditions.
These findings, combined with previous high-field \usr, NMR, STM and transport studies, demonstrate that the charge and spin channels are strongly intertwined in these materials.

Furthermore, this study highlights the effectiveness of ALC \usr measurements as a powerful tool for probing electronic orders. Additional experimental and computational investigations, potentially utilizing resonances of other quadrupolar nuclei (Rb, Sn, Cs, K), will be essential to further elucidate the microscopic long-range order below $T^{*}$ in this and related compounds.

\section{Acknowledgments}
We thank Roberto De Renzi and Giuseppe Allodi for insightful discussions.
The computational resources were provided by the SCARF cluster of the STFC Scientific Computing Department and by the ISCRA initiative of CINECA with project IsCb6\_TRSBKS.
Work in Parma was funded by the PNRR MUR project ECS-00000033-ECOSISTER.
IO acknowledges support by University of Parma through the action Bando di Ateneo 2023 per la ricerca.
SDW and ACS gratefully acknowledge support via the UC Santa Barbara NSF Quantum Foundry funded via the Q-AMASE-i program under award DMR-1906325.

\bibliographystyle{apsrev4-2}
\bibliography{main}

\onecolumngrid{}

\supplementarysection{}

\pagebreak
\section*{Supplementary Information to ``Unveiling the nature of electronic transitions in RbV$_3$Sb$_5$ with Avoided Level Crossing $\mu$SR''}
\section{Analysis of EFG obtained from plane-wave based simulations}
\begin{figure}
    \centering
    \includegraphics[width=0.45\linewidth]{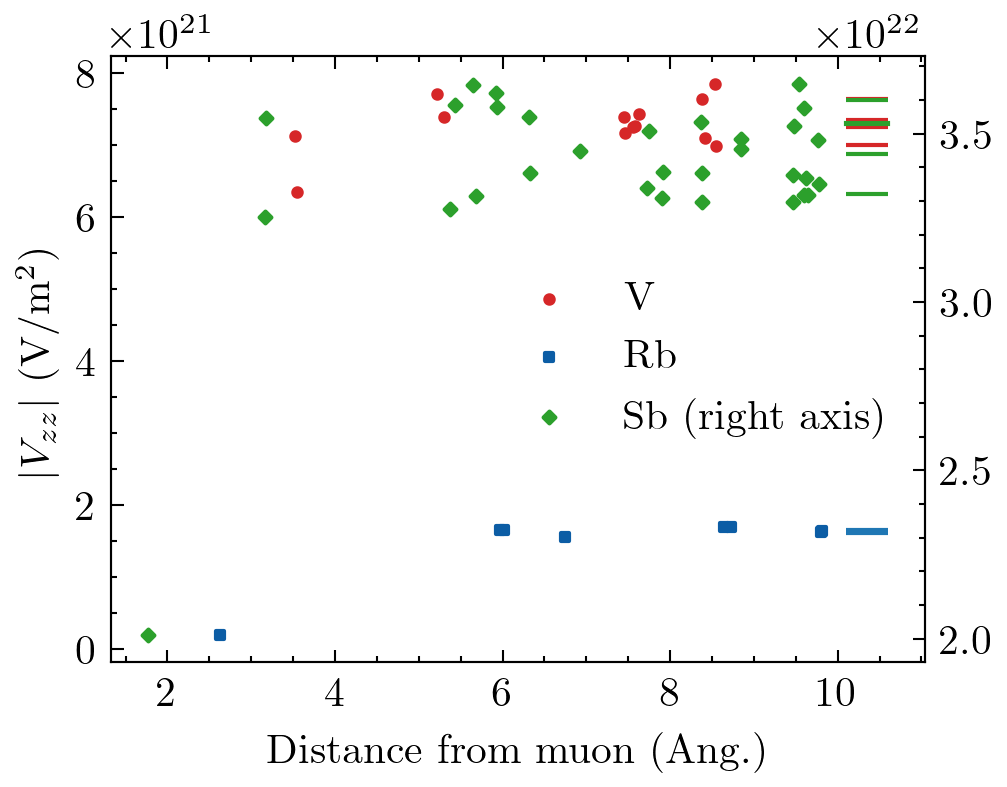}
    \includegraphics[width=0.45\linewidth]{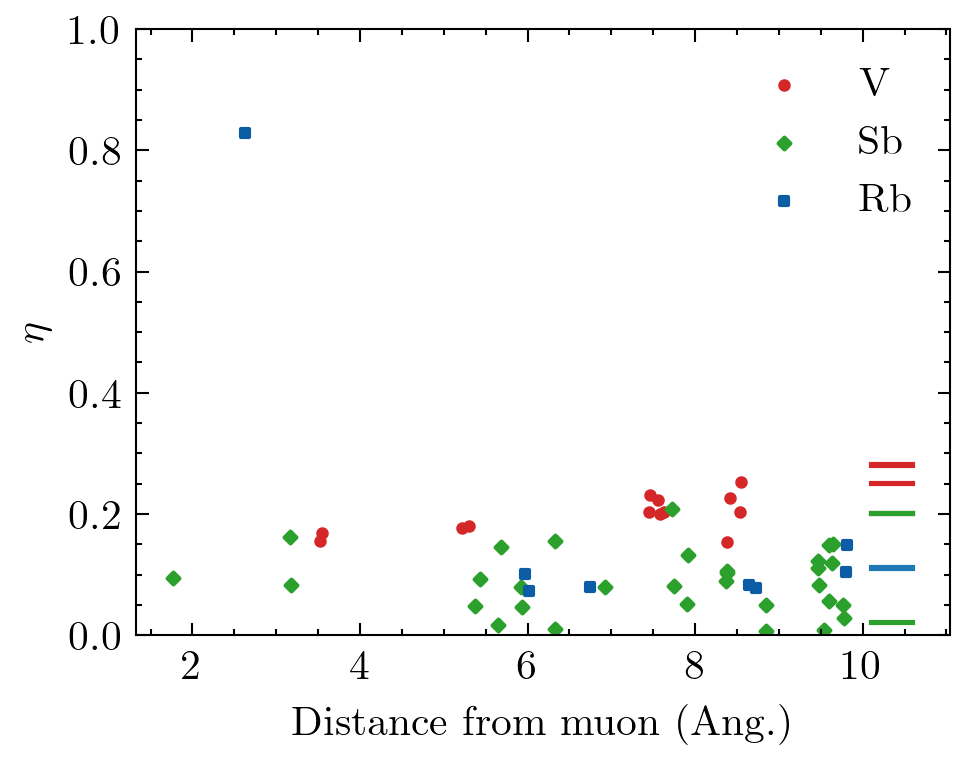}
    \caption{The values of $V_{zz}$ (left) and of the asymmetry parameter $\eta$ (right) for each atom in the PW-based supercell simulation of Ref.~\cite{materialscloud} reported as a function of the distance from the muon site. The values for the unperturbed lattice in the tri-hexagonal phase are reported as colored thick bars on the right side of each plot.}
    \label{fig:Vzz}
\end{figure}
The embedding position of the positive muon and the perturbation that it induces on the Electric Field Gradient (EFG) of its neighboring atoms is obtained by analyzing the computational results published in Ref.~\cite{Graham_2024}.
The description of the local distortion of the lattice can be found in the supplemental information of the cited reference, the perturbation of the EFG at the nuclear sites is
shown in Fig.~\ref{fig:Vzz}. As shown in the figure, $V_{zz}$ (the largest component of the EFG in the principal axis system) of the nearest neighbor (NN) Sb atom is strongly perturbed by the positive interstitial charge of the muon. The second NN, Rb, also shows a substantial reduction of $V_{zz}$ while V atoms (red points) are almost unperturbed, with a small decrease of $V_{zz}$ by less than 10 \%. The asymmetry parameter $\eta$ of the nearest Rb atom shows the largest deviation from the unperturbed value. At large distance, both $V_{zz}$ and $\eta$ for the various atoms reach the unperturbed values indicated by the dashed line in the figure (obtained with additional simulations using the same code and computational parameters reported in Ref.~\cite{Graham_2024}).

The most accurate results for the EFG at V sites are  obtained with a full-potential approach using augmented plane waves. Here we report the estimates provided in Ref.\cite{Frassineti_materialscloud, Frassineti_2023} in the last two columns of Tab.~\ref{tab:efg-from-apw} which are used to produce Fig.~4a in the main text. Comparison can be made with the unperturbed values obtained for the same lattice structure with a PW based method (second and third columns of Tab.~\ref{tab:efg-from-apw}) 

\begin{table}[h]
    \centering
    \begin{tabular}{c|c|c|c|c}
     & \multicolumn{2}{c|}{PW} & \multicolumn{2}{c}{APW} \\
    \hline
    Atom & $|V_{zz}|$ ($10^{22}$ Vm$^{-2}$) & $\eta$ & $|V_{zz}|$ ($10^{22}$ Vm$^{-2}$) & $\eta$\\
    \hline
    Sb & 3.34 & 0.065 & 3.31 & 0.111 \\
    Sb & 3.49 & 0.0 & 3.52 & 0.0 \\
    Sb & 3.05 & 0.039 & 3.20 & 0.058 \\
    Sb & 3.10 & 0.0 & 3.32 & 0.0 \\
    Rb & 0.137 & 0 & 0.093 & 0.0 \\
    Rb & 0.140 & 0.024 & 0.110 & 0.127 \\
    V & 0.693 & 0.241 & 0.660 & 0.431 \\
    V & 0.731 & 0.221 & 0.697 & 0.432
    \end{tabular}
    \caption{Value of EFG parameters for the inequivalent nuclei in the Tri-Hexagonal structure as obtained from APW simulations.}
    \label{tab:efg-from-apw}
\end{table}

\section{Predicted resonances and fit of ALC resonances}
\begin{figure}
    \centering
    \includegraphics[width=0.5\linewidth]{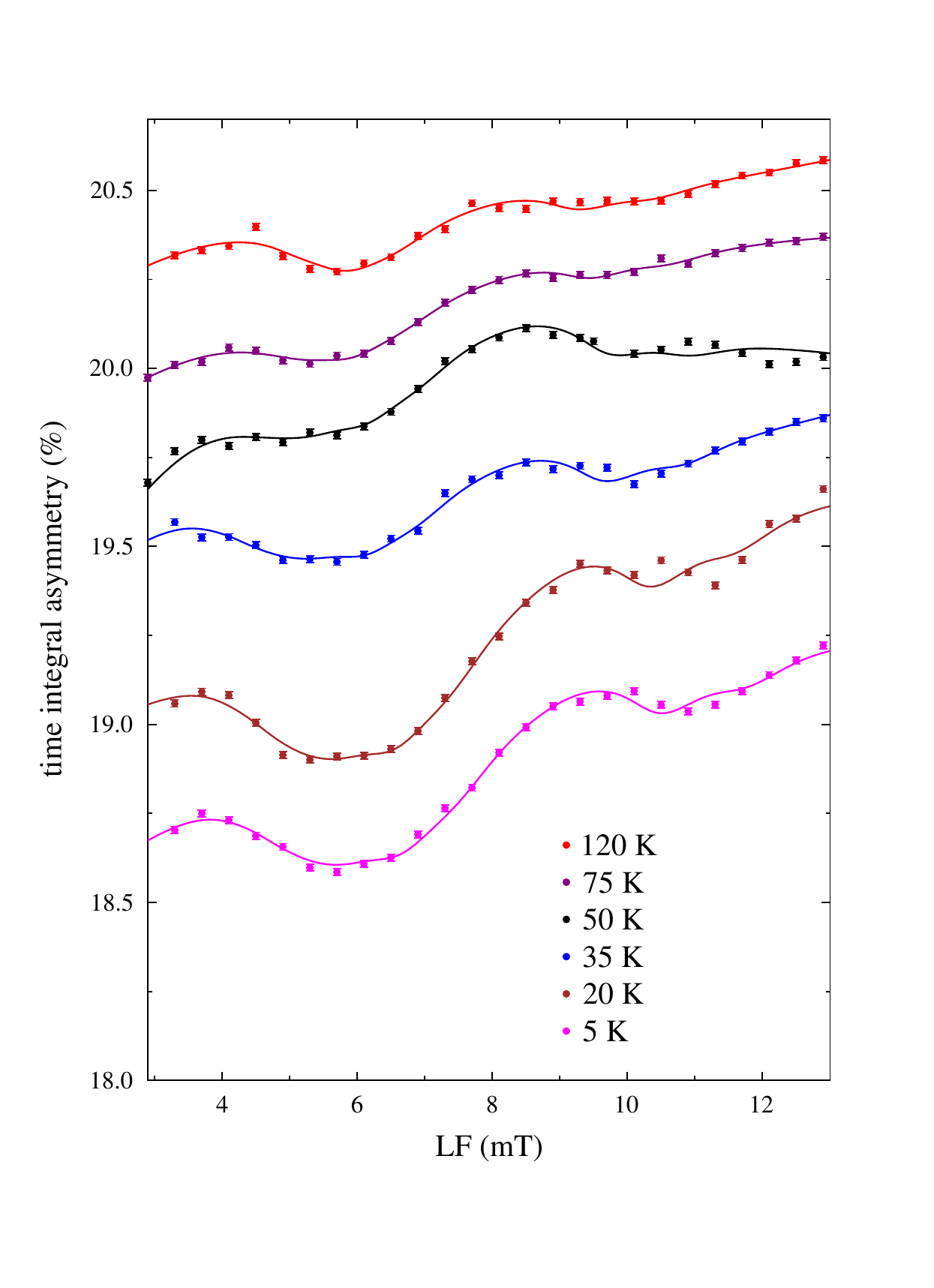}
    \caption{Fit of time integrated asymmetry according to the model reported in the main text. Scans are displaced vertically for clarity.}
    \label{fig:Vfits}
\end{figure}

\begin{figure}
    \centering
    \includegraphics[width=0.7\linewidth]{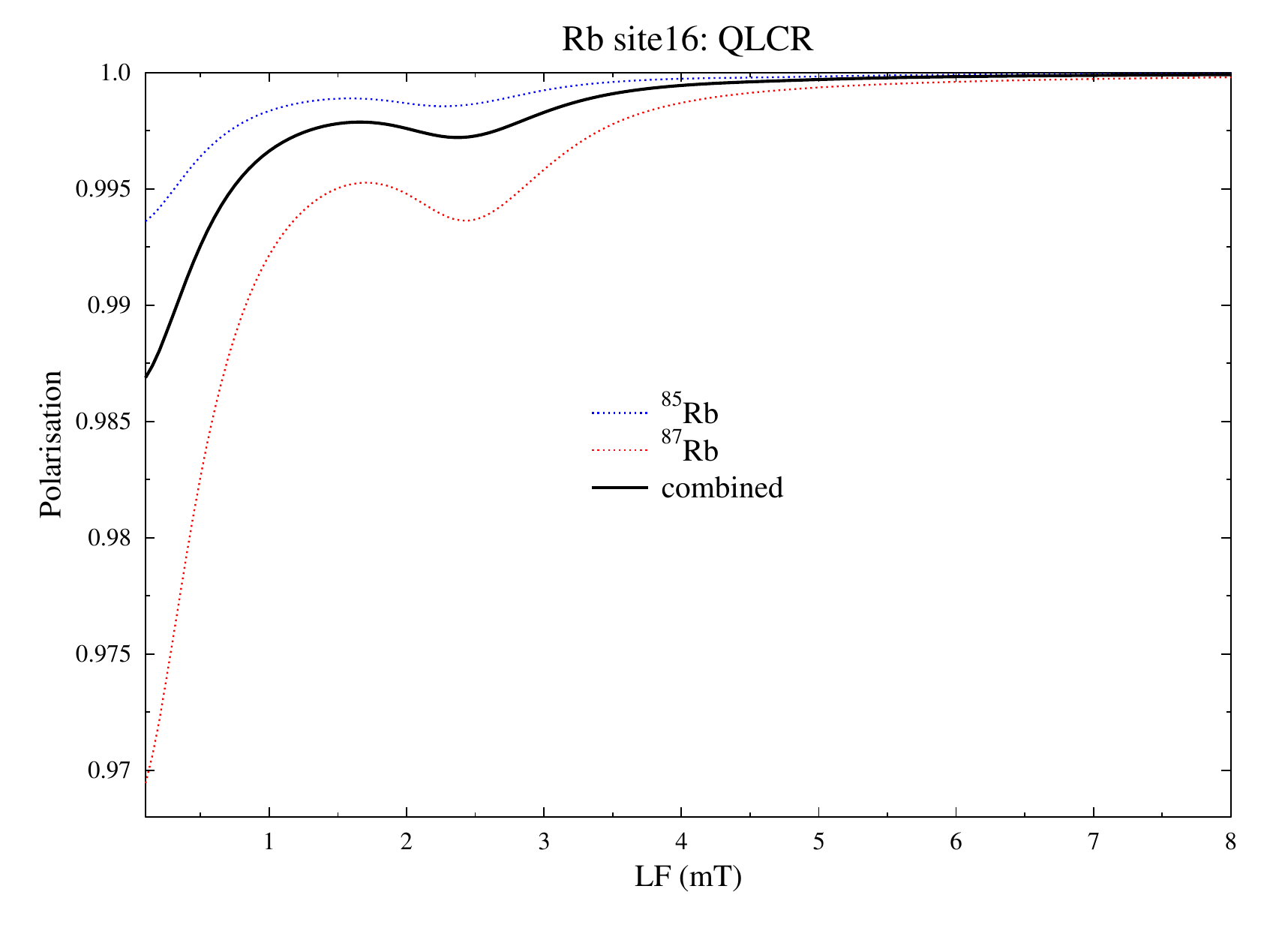}
    \caption{Resonances predicted for Rb nuclei using the EFG tensors from PW-DFT simulations.}
    \label{fig:Rb}
\end{figure}

\begin{figure}
    \centering
\includegraphics[width=0.7\linewidth]{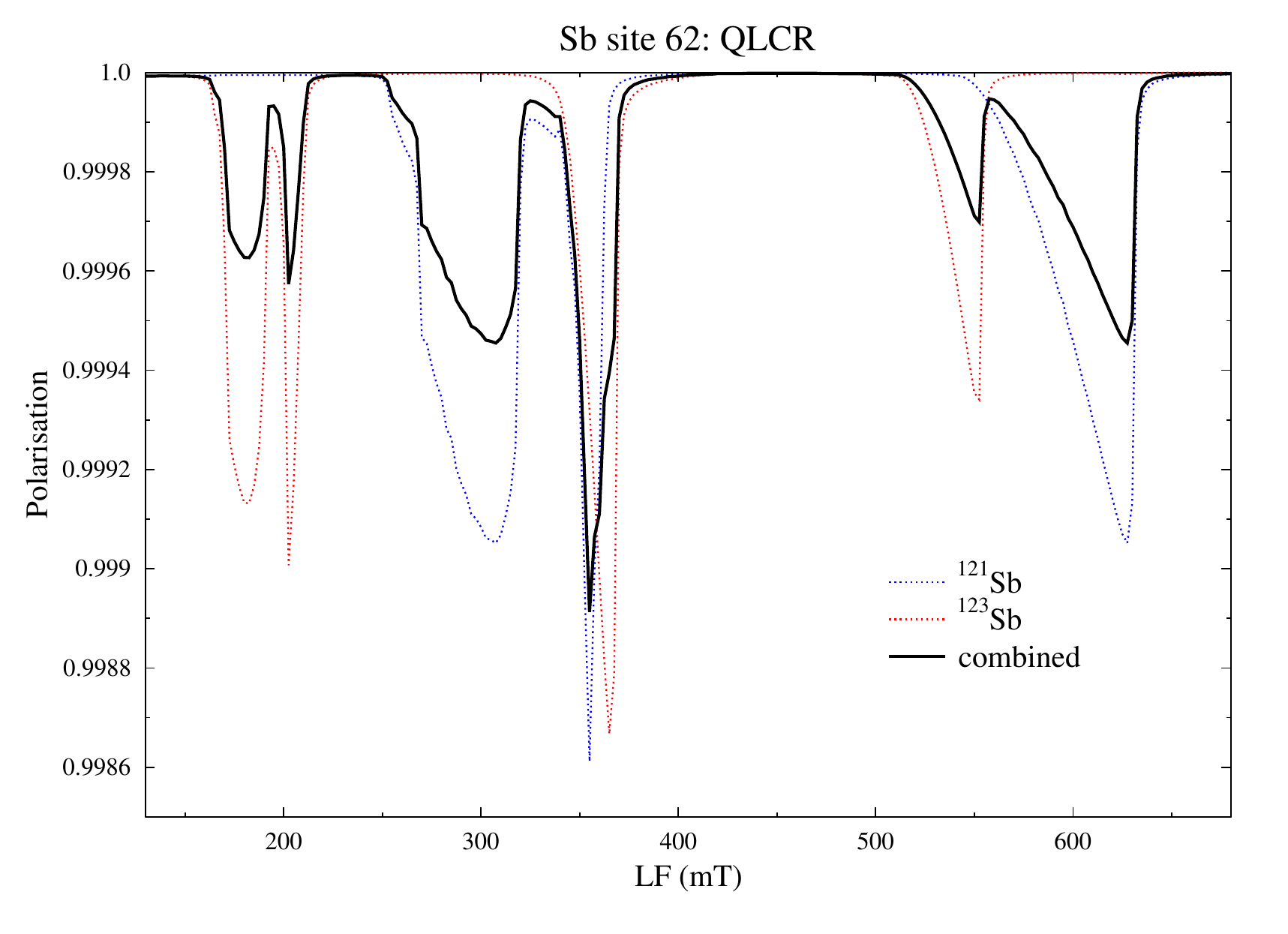}
    \caption{Resonances predicted for Sb nuclei using the EFG tensors from PW-DFT simulations.}
    \label{fig:Sb}
\end{figure}

The ALC resonances are computed with two different approaches.
In a first analysis, the interaction between the muon an a single nucleus is considered and the resulting
 asymmetry is integrated over the range 0 to 24 $\mu$s.
The results obtained for the NN Rb and Sb nuclei of the muon are shown in Fig.~\ref{fig:Rb} and Fig.~\ref{fig:Sb}
respectively. From these it can be seen that the Rb QLCR spectrum is weak and at very low field and the Sb spectrum is also weak and at much higher field than our measurement range.
Having ruled out contributions from these nuclei, the same approach is used in the main text to fit the experimental results considering only the contribution of V atoms. The resulting best fits are shown in Fig.~\ref{fig:Vfits} with corresponding values reported in Tab.~\ref{tab:fitdata} and plotted in Fig.4b of the main text.

A second set of simulations is performed with the method proposed by Celio~\cite{Celio_1986}. In this case we consider a larger Hilbert space which describes 6 V atoms and, in one case, also the nearest neighbor Sb atom of the muon.
The powder average is performed using 500 random directions. The result is shown in Fig.~4a in the main text.


\begin{table}[]
    \centering
    \begin{tabular}{c|c|c|c|c|c|c|c|c}
T &     NQCC factor (Err) & $\eta$ (Err)  &        Amplitude (Err)  &  A0 (Err)  &         BG-lin (Err)   &    BG-quad (Err)    &   BG-cubic (Err) & $\chi^2$ \\
\hline

120 & 0.9034(62) & 0.629(16) & 0.852(50) & 19.741(34) & 0.0272(17) & -0.000306(23) & 0.00000113(9) &  4.439  \\
 75 & 0.9629(62) & 0.451(14) & 0.735(58) & 19.291(30) & 0.0184(17) & -0.000153(23) & 0.00000045(9) &  2.060 \\
 50 & 0.9871(37) & 0.3889(56) & 1.353(61) & 19.277(29) & 0.0421(16) & -0.000418(21) & 0.00000126(9) &  13.419 \\
 35 & 0.9826(38) & 0.4020(63) & 1.356(56) & 19.432(30) & 0.0339(16) & -0.000376(22) & 0.00000138(9) &  7.989 \\
 20 & 1.0525(24) & 0.3919(36) & 2.176(63) & 19.532(31) & 0.0202(17) & -0.000135(23) & 0.00000033(9) & 15.135 \\
  5 & 1.0685(26) & 0.3928(39) & 2.075(64) & 19.283(31) & 0.0290(17) & -0.000245(23) & 0.00000072(9) &  9.133  \\   
    \end{tabular}
    \caption{Results of the fit shown in Fig.~\ref{fig:Vfits} and described in the main text. }
    \label{tab:fitdata}
\end{table}

\section{Additional DFT simulations}
The values for the unperturbed EFG in the PW basis reported in Tab.~\ref{tab:efg-from-apw} are obtained using augmented plane wave pseudopotentials of the PSlibrary \cite{DalCorso_2014}. The PBEsol \cite{PhysRevLett.100.136406} functional has been used to approximate the exchange and correlation contribution and the reciprocal space has been sampled with a $8 \times 8 \times 8$ Monkhorst-Pack  grid \cite{PhysRevB.13.5188}. The plane wave cutoff is set to 90~Ry while charge density is expanded up to 1020~Ry. The EFG tensors are obtained with the GIPAW code \cite{gipaw}.

\end{document}